\documentclass[
aps,prl,twocolumn,10pt,floatfix,showpacs,preprintnumbers,
  amsmath,amssymb,
  nobibnotes
]{revtex4-2}

\usepackage{graphicx}
\usepackage{dcolumn}
\usepackage{bm}
\usepackage{wrapfig}
\usepackage{graphicx}
\usepackage{xcolor}
\usepackage{amsmath}
\usepackage{amssymb}
\usepackage{amsfonts}
\usepackage{mathrsfs}
\usepackage{latexsym}
\usepackage{relsize}
\usepackage{dcolumn}
\usepackage{bm}
\usepackage{url,lineno}
\usepackage{color}
\usepackage[english]{babel}

\renewcommand{\le}{\leqslant}

\newcommand{\be}{\begin{equation}}
\newcommand{\en}{\end{equation}}

\renewcommand{\vec}[1]{\boldsymbol{#1}}

\usepackage{cuted}
\begin{document}

\title{Supercontraction-Induced Twist in Spider Silk Is a Dual Poynting Effect}
\date{\today}

\author{V. Fazio}
\affiliation{Dipartimento di Scienze dell'Ingegneria Civile e dell'Architettura,
Politecnico di Bari, Italy}

\author{G. Puglisi}
\email{giuseppe.puglisi@poliba.it}
\affiliation{Dipartimento di Scienze dell'Ingegneria Civile e dell'Architettura,
Politecnico di Bari, Italy}

\author{G. Saccomandi}
\affiliation{Dipartimento di Ingegneria Industriale, Universit\`a degli Studi di Perugia, Italy}
\affiliation{School of Mathematics, Statistics and Applied Mathematics,
University of Galway, University Road, Galway, Ireland}

\date{}
\begin{abstract}
\noindent
Spider dragline silk supercontracts as humidity increases, displaying large axial shortening together with a reproducible macroscopic twist. The physical origin of this torsion remains debated and is often attributed to helically arranged load-bearing elements, despite the lack of direct evidence for helicity in the native fiber.
Here we show that torsion can arise generically from nonlinear anisotropic elasticity: humidity-driven shortening of the amorphous matrix, mechanically constrained by stiff, axially aligned $\beta$-sheet--rich load-bearing segments and their experimentally induced prestretch, drives the system into a dual Poynting regime in which axial shortening couples to spontaneous twist.
Coupling a diffusion-based water-uptake law to irreversible matrix remodeling and fiber plasticity, the model quantitatively reproduces monotonic and cyclic torsional measurements using parameter values consistent with available experimental material parameters.
These results identify supercontraction-induced torsion in spider silk as a manifestation of a dual Poynting effect and provide a minimal, physically grounded framework for humidity-driven torsional actuation in matrix--fiber architectures.
\end{abstract}

\maketitle

\noindent

\noindent

Spider dragline silk displays exceptional strength and toughness arising from its efficient hierarchical protein architecture \cite{BP,Nova}. Despite intensive efforts to produce artificial spider silks and silk-inspired fibers, the outstanding performance of natural dragline silk remains largely unattained \cite{johansson2021doing}. This calls for physically based multiscale models capable of identifying the key physical ingredients that control the macroscopic response.

Motivated by both biological function and technological applications, the influence of environmental conditions has been extensively investigated, revealing the central role of temperature and humidity in the mechanical response of silk fibers \cite{perez2021basic,Faz3}.
In particular, above a critical relative humidity ($RH \sim 70 \%$), the fiber undergoes supercontraction \cite{sca,scb,scc,Bue}, a pronounced axial shortening that can reach 40--60$\%$, driven by humidity-induced changes in its semicrystalline polymer structure \cite{cohen2021origin,JMPSNoi}.
This humidity-driven contraction gives spider silk actuation performance that is competitive with state-of-the-art humidity-responsive materials and, by standard efficiency metrics, compares favorably with many other actuator classes \cite{Faz1}.
Beyond axial shortening, recent experiments have revealed that supercontraction can be accompanied by a sizable macroscopic torsion of the fiber \cite{expexp}. Torsional actuation is comparatively rare, and the physical origin of this twist remains unresolved.

Explanations based on helically arranged load-bearing fibrils \cite{Co} remain speculative: helical motifs are observed to emerge predominantly after supercontraction \cite{Vol}, while high-resolution imaging of native dragline silk provides no conclusive evidence for a helical reinforcement architecture \cite{Xu}. Moreover, under dry conditions, recent experiments indicate that the load-bearing fibrils are essentially parallel to the fiber axis \cite{jenkins2013characterizing,wang2018strength,perera2023natural}.

This raises a fundamental question: can humidity-induced torsion arise solely from nonlinear anisotropic elasticity, without invoking intrinsic geometric chirality of the microstructure?

To address this question, we build on the nonlinear theory of fiber-reinforced solids and on the recently identified dual Poynting effect \cite{FPS}.
In classical nonlinear elasticity, applying a torque to an incompressible cylinder inevitably induces an axial length change,  known as the Poynting effect\cite{Poynting,Zurlo,HM}.
We recently showed that the converse also holds for fiber-reinforced materials: an imposed axial load can generate a spontaneous torsion --the dual Poynting effect-- when sufficiently stiff fibers are aligned with the cylinder axis \cite{FPS}.

Guided by the dual Poynting effect, we show that humidity-induced matrix shortening, constrained by stiff coaxial $\beta$-sheet--rich crystalline segments \cite{Jen,hd,JMPSNoi,expexp}, provides the internal kinematic mismatch that plays the role of the imposed compression in Ref.~\cite{FPS}, thereby coupling axial remodeling to torsion. 
We formulate a microstructure-inspired fiber-reinforced model with matrix remodeling and an evolving natural configuration (residual stretch) of the load-bearing phase, and we quantitatively reproduce the measured twist in monotonic and cyclic humidity protocols \cite{expexp} without prescribing microscopic helicity.


\paragraph{Wire model---}
Motivated by molecular and structural evidence that dragline silk mechanics is governed by axially oriented load-bearing protein segments with crystalline $\beta$-sheet nanodomains embedded in an amorphous matrix, we model the fiber as an incompressible fiber-reinforced cylinder with $\beta$-sheet–rich load-bearing segments with the fiber axis \cite{vanBeek2002molecular,Sirichaisit2000,Glisovic2008}.

Within classical nonlinear elasticity for incompressible fiber-reinforced continua \cite{Holzapfel2010}, we describe the constitutive response through a strain-energy density
\[
W=W(I_1,I_4,I_5),
\]
where $\vec{F}$ is the deformation gradient, $\vec B= \vec F\vec F^{T}$ the left Cauchy--Green deformation tensor, and $I_1=\mathrm{tr}(\vec B)$. Denoting by $\vec A$ the unit fiber direction in the reference configuration and by $\vec a=\vec F\vec A$ the deformed fiber vector, we introduce the standard fiber invariants
\[
I_4=\vec a\cdot\vec a,\qquad
I_5=\vec a\cdot(\vec B \vec a).
\]
Here $I_4=\lambda_f^2$ measures the squared stretch of the fiber family along $\vec A$, whereas $I_5$ accounts for matrix--fiber interaction through shear (fiber-directional distortion) in the deformed configuration (see, e.g., \cite{Holzapfel2010}).

Since the fibers are assumed to be aligned with the (fiber) $Z$ direction, we set $\vec{A}\equiv \vec{E}_3$. Using cylindrical coordinates $(R,\Theta,Z)$ in the reference configuration and $(r,\theta,z)$ in the current configuration, we consider the class of extension-torsion deformations
\be \label{eq1}
r=\lambda^{-1/2}R, \qquad \theta=\Theta+\kappa\,\lambda Z, \qquad z=\lambda Z
\en
where $\lambda$ is the axial stretch and $\kappa$ the twist per unit present length.
\begin{figure}[t]
\includegraphics[width=0.48\textwidth]{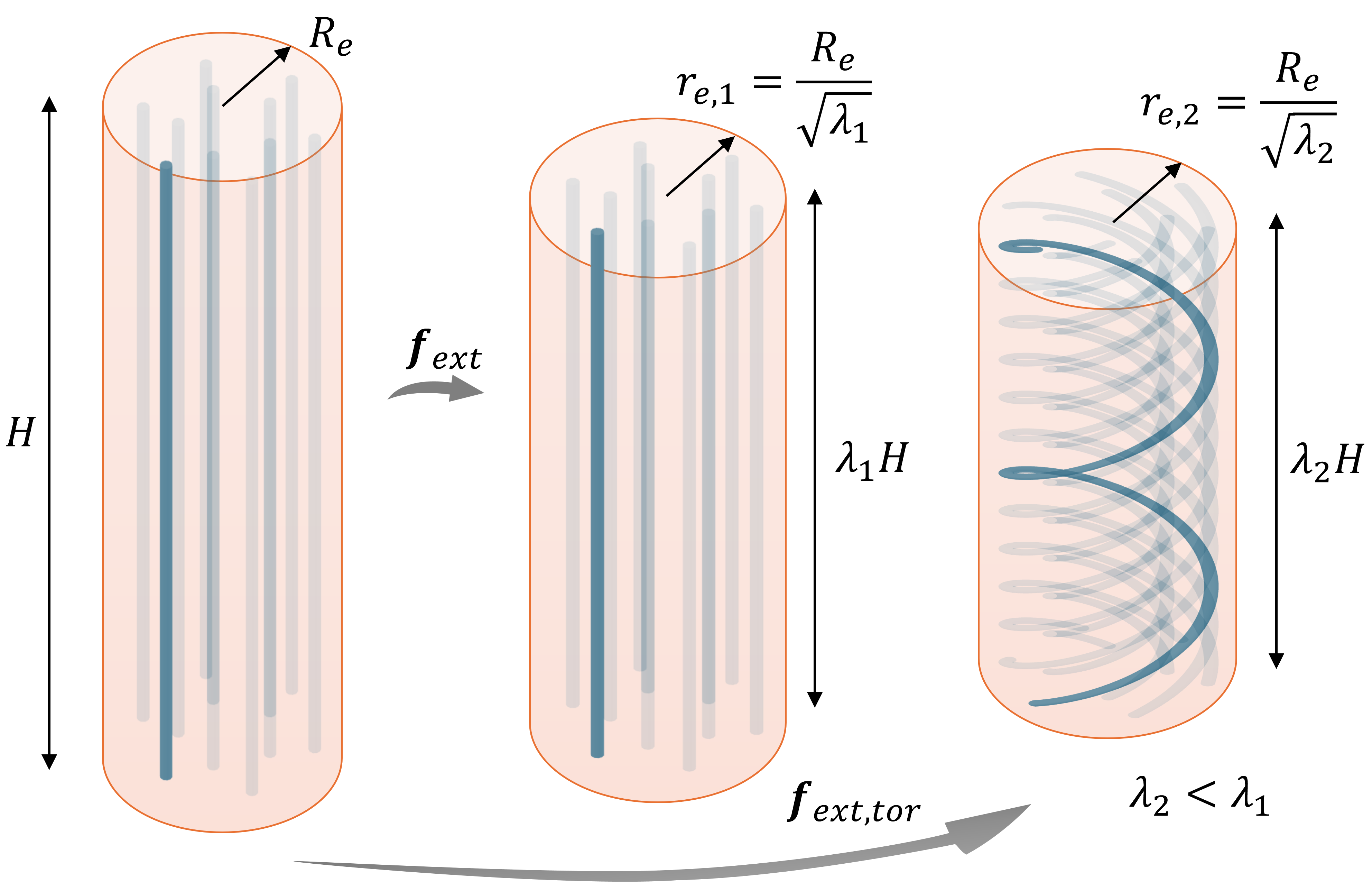}
\caption{Scheme of spider silk structure and wetting supercontraction effect. In the middle scheme we show the classical purely extensional assumption and in the third one we show the experimentally observed combined extensional-torsional supercontraction behavior}
\label{sca}
\end{figure}

\noindent Introducing the reference radius $R_e$ of the cylinder, we define the dimensionless radial coordinate $\bar R:=R/R_e$ and the dimensionless twist $\bar\kappa:=\kappa R_e$. With these definitions, the left Cauchy--Green deformation tensor reads
\[
[\vec{B}]=
\begin{pmatrix}
	\lambda^{-1} & 0 & 0 \\
	0 & \lambda^{-1}+\lambda\,\bar R^2 \bar\kappa^2 & \lambda^{3/2}\bar R\bar\kappa \\
	0 & \lambda^{3/2}\bar R\bar\kappa & \lambda^2
\end{pmatrix},
\]
and the corresponding invariants take the form
\begin{equation}
\begin{aligned}
I_1 &= \frac{2}{\lambda} + \lambda^2 + \lambda\,\bar\kappa^2 \bar R^2, \\
I_4 &= \lambda^2 + \lambda\,\bar\kappa^2 \bar R^2, \\
I_5 &= \bar R^2 \bar\kappa^2 + \lambda^2\bigl(\lambda + \bar R^2 \bar\kappa^2\bigr)^2.
\end{aligned}
\end{equation}

The Cauchy stress reads \cite{Holzapfel2010}
\[
\begin{array}{llll}
	\vec{T} =	- p\,\vec{I} + 2 W_1 \vec{B} 
	+ 2 W_4 (\vec{F}\vec{A} \otimes \vec{F}\vec{A}) \\ \vspace{.3cm}
	+ 2 W_5 \left[ (\vec{F}\vec{A} \otimes \vec{B}\vec{F}\vec{A}) + (\vec{B}\vec{F}\vec{A} \otimes \vec{F}\vec{A}) \right],
\end{array}
\]
where $W_i := \partial W / \partial I_i$ and $p$ is the Lagrange multiplier enforcing incompressibility.

Following Horgan and Murphy~\cite{hm2020}, and neglecting for simplicity the $I_2$ dependence, we adopt a transversely isotropic strain-energy density of the form
\be
\begin{array}{lll}
	W = \frac{\mu_T}{2}(I_1 - 3) + (\mu_T - \mu_L)(X - X_0) \\[4pt]
	\qquad\qquad
	+ \frac{E_L + \mu_T - 4 \mu_L}{8} (X - X_0)^2
	+ \frac{\mu_L - \mu_T}{2} (I_5 - 1),
\end{array} \label{stene}
\en
with 
\begin{equation}\label{eq:XY_v3}
X = I_4 + \alpha I_1,\qquad
\end{equation}
where $\alpha>0$ is a dimensionless coupling parameter and
$X_0 = 1 + 3\alpha$.
The constants $\mu_T$, $\mu_L$, and $E_L$ denote the transverse and longitudinal shear moduli and the longitudinal Young modulus, respectively. In the present setting, torsion is driven by internal remodeling rather than an imposed axial compression; an explicit coupling between axial fibers and matrix shear, captured by the $Y$-term, is therefore essential to activate the torsional response. More elaborate constitutive couplings are possible, but the structure~\eqref{eq:XY_v3} is sufficient for the purposes of the present work. Accordingly, in the present setting, the dependence on the second isotropic invariant $I_2$ is neglected, as it does not play a role in the torsional response.
 The form~\eqref{stene} provides a minimal yet effective representation of fiber--matrix interaction, consistent with that employed to investigate the role of such coupling in the Poynting effect for fibrous soft biomaterials under torsion~\cite{horgan2021effect}.
This choice ensures vanishing energy and deviatoric stress in the undeformed configuration and, upon linearization, recovers classical incompressible transversely isotropic elasticity~\cite{hm2020} (see End Matter).

\begin{figure*}[t] 
\centering
\includegraphics[width=.85\textwidth]{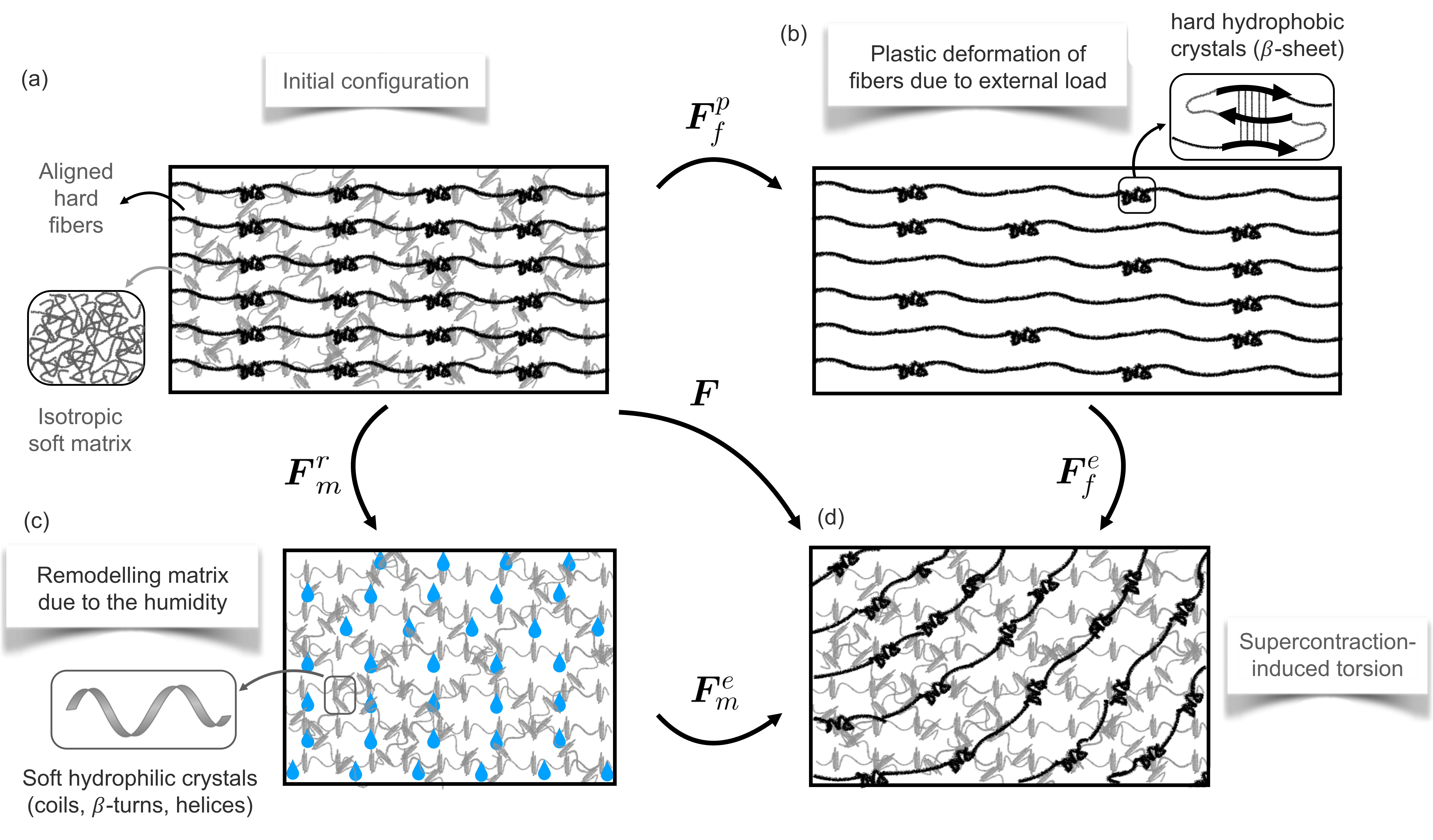}
\caption{\label{sc}Microstructure-inspired schematic of the constitutive assumptions. 
(a) In the dry reference state, aligned stiff protein chains associated with hydrophobic $\beta$-sheet nanocrystals act as reinforcing fibers embedded in an isotropic soft amorphous matrix. 
(b) Under an external axial load, these reinforcing fibers undergo plastic deformation, resulting in a residual prestretch. 
(c) Upon hydration, water molecules disrupt hydrogen bonds in hydrophilic secondary structures (coils, $\beta$-turns, helices) within the amorphous matrix, enabling entropic recoiling of the chains and an effective shortening of the matrix. 
(d) In the supercontracted state, this matrix shortening, transmitted through matrix--fiber coupling to the prestretched crystalline fraction, gives rise to both macroscopic axial shortening and torsional deformation of the silk fiber.}
\end{figure*}

In the context of spider silk, we interpret this constitutive law as describing a composite material comprising two main fractions: a crystalline, hydrophobic fraction, dominated by $\beta$-sheet nanodomains that are largely unaffected by water molecules, and an amorphous, hydrophilic fraction, rich in $\beta$-turns and helical structures, which undergoes humidity-driven remodeling and is responsible for supercontraction (Fig.~\ref{sc}). \vspace{0.3 cm}

\paragraph*{Theoretical description of humidity-induced torsion---}
In the torsional experiments reported in Ref.~\cite{expexp}, a small paddle is attached to the silk fiber in order to straighten the initially curved fiber. At the stress levels reached under these conditions, based on experimental evidence described in the following, the crystalline, hydrophobic fraction of spider silk undergoes irreversible molecular rearrangements associated with partial unraveling of $\beta$-sheet structures \cite{JMPSNoi}. Upon unloading, the reinforcing fibers therefore retain a residual prestretch, which we model through a `plastic-type'
deformation gradient
\begin{align*}
	\vec{F}^p_f &= \lambda_f^{-1/2} \bigl( \vec{I} - \vec{E}_3 \otimes \vec{E}_3 \bigr)
	+ \lambda_f\, \vec{E}_3 \otimes \vec{E}_3,
\end{align*}
where $\lambda_f=\lambda_f(\lambda_{\max})$ denotes the residual fiber stretch as a function of the maximum stretch $\lambda_{\max}$ attained during loading.

Hydration primarily affects the amorphous, hydrophilic fraction of the silk. In the dry state, the amorphous protein chains are effectively `frozen' in elongated configurations by a dense network of hydrogen bonds. Water uptake disrupts these bonds, allowing entropic recoiling of the chains and leading to supercontraction. We describe this humidity-driven matrix remodeling by a deformation gradient
\begin{equation} \label{eqx}
	\vec{F}^r_m 
	= \lambda_m^{-1/2} \bigl( \vec{I} - \vec{E}_3 \otimes \vec{E}_3 \bigr)
	+ \lambda_m\, \vec{E}_3 \otimes \vec{E}_3,
\end{equation}
where $\lambda_m=\lambda_m(c)<1$ is the stretch of the matrix along the fiber axis depending on the water volume fraction $c=c(RH)$ in the fiber, where $RH$ is  the relative humidity. The transverse expansion implied by~\eqref{eqx} ensures preservation of the matrix volume in the remodeled configuration, consistent with the essentially isochoric nature of supercontraction \cite{guinea2005stretching}.

As a result, humidity-induced shortening of the amorphous matrix occurs relative to a prestretched crystalline network, so that the remodeled configuration does not coincide with the original stress-free reference state. This internal mismatch of fibers and matrix references configuration provides a driving mechanism analogous to that underlying the classical dual Poynting effect \cite{FPS} where the bifurcation is induced by an imposed cylinder shortening or assigned compressive force. Consequently, under supercontraction, a spontaneous torsional deformation emerges, even in the presence of an externally applied {\it tensile} load. The observed torsion is thus described not a consequence of geometric chirality or helical microstructure, but rather as the manifestation of a dual Poynting-type mechanism driven by internal remodeling. \vspace{0.3 cm}

\paragraph*{Kinematic compatibility and overall response---}
After supercontraction, the spider silk composite undergoes an overall deformation $\vec f$, with $\vec F=\nabla\vec f$ denoting the associated deformation gradient, mapping the dry, virgin configuration into the supercontracted wet state (Fig.~\ref{sc}(d)). This macroscopic deformation must satisfy kinematic compatibility between the amorphous matrix and the crystalline reinforcing fibers, ensuring that both material phases attain a common final configuration despite having undergone distinct remodeling processes. As a result, the elastic response of each phase is measured relative to its own natural configuration, whereas the observable deformation $\vec F$ enforces compatibility in the supercontracted state.

We denote the elastic deformation gradients of the amorphous matrix and of the reinforcing fibers by
\[
\vec{F}^e_m=\vec{F}(\vec{F}^r_m)^{-1},\qquad
\vec{F}^e_f=\vec{F}(\vec{F}^p_f)^{-1}.
\]
The associated left Cauchy--Green tensors are
\[
\vec{B}^e_m=\vec{F}^e_m(\vec{F}^e_m)^T,\qquad
\vec{B}^e_f=\vec{F}^e_f(\vec{F}^e_f)^T,
\]
and the relevant elastic invariants read
\[
I_1^m=\mathrm{tr}\,\vec{B}^e_m,\qquad
I_4^f=\vec a_f\cdot\vec a_f,\qquad
I_5^f=\vec a_f\cdot(\vec{B}^e_f\,\vec a_f),
\]
with $\vec a_f=\vec{F}^e_f\vec{A}$ and $\vec{A}$ denoting the unit vector identifying the fiber direction in the reference configuration.

Since the matrix and fibers deform elastically relative to their respective remodeled and plastic reference configurations, the strain--energy density \eqref{stene} is written in terms of the corresponding elastic invariants as
\begin{equation}
\begin{aligned}
	W&=
	\tfrac{\mu_T}{2}(I_1^m-3)
	+(\mu_T-\mu_L)\bigl[(I_4^f+\alpha I_1^m)-(1+3\alpha)\bigr]\\
	&\quad
	+\tfrac{\mu_L-\mu_T}{2}\bigl[I_5^f-(1+6\alpha)\bigr]
	+\tfrac{E_L+\mu_T-4\mu_L}{8}\bigl[(I_4^f+\alpha I_1^m)\\
	&\quad-(1+3\alpha)\bigr]^2.
\end{aligned}
\label{eq:Wlocal}
\end{equation}

The resulting Cauchy stress reads
$$
\begin{aligned}
	\vec{T} &=
	-\,p(R)\,\vec{I}
	+2W_{1}^m\,\vec{B}^e_m
	+2W_{4}^f\,(\vec{F}^e_f\vec{A}\otimes\vec{F}^e_f\vec{A})\\
	&\quad
	+2W_{5}^f\Big[\vec{F}^e_f\vec{A}\otimes(\vec{B}^e_f\vec{F}^e_f\vec{A})
	+(\vec{B}^e_f\vec{F}^e_f\vec{A})\otimes\vec{F}^e_f\vec{A}\Big],
\end{aligned}
$$
where $p(R)$ enforces incompressibility and  $W_{1}^m=\partial W/\partial I_1^m$, $W_{4}^f=\partial W/\partial I_4^f$, and $W_{5}^f=\partial W/\partial I_5^f$.

The deformation~\eqref{eq1} belongs to a universal class for incompressible cylinders and therefore admits a pressure field $p=p(R)$ such that the equilibrium condition $\mathrm{div}\,\vec{T}=\vec{0}$ is satisfied. Imposing the traction-free boundary condition on the lateral surface, $\vec{T}\vec{e}_r\cdot\vec{e}_r=0$, the pressure $p(R)$ can be determined explicitly.

The corresponding nondimensional torque associated with the deformation~\eqref{eq1} is then given by
\be \label{tor}
\bar{M}
= \frac{1}{\mu_T \pi R_e^{3}}\,
2\pi \int_{0}^{R_e}
\frac{T_{32}\, R^{2}}{\lambda^{3/2}}\, dR .
\en
\noindent and for the non dimensional axial force 
\be 
\label{force} 
\bar N=
\frac{1}{\mu_T \pi R_e^2}\,
2\pi \int_{0}^{R_e} \frac{T_{33}\, R}{\lambda}\, dR.
\en 

If a torque is prescribed under vanishing resultant axial force, $\bar N=0$ and $\bar M\neq0$, Eqs.~\eqref{tor}--\eqref{force} imply that torsion cannot occur at fixed length: any twist ($\bar\kappa\neq0$) necessarily induces $\lambda\neq1$, recovering the classical Poynting effect for incompressible cylinders.

Conversely, if an axial force is prescribed under zero torque, $\bar N\neq0$ and $\bar M=0$, a trivial equilibrium branch with $\kappa=0$ always exists. For isotropic materials this branch is unique, whereas for fiber-reinforced solids the condition $\bar M(\lambda,\kappa)=0$ may admit an additional torsional solution $\kappa=\hat\kappa(\lambda)$. This nontrivial branch corresponds to the dual Poynting regime identified in Ref.~\cite{FPS}. We emphasize that $\lambda$ is the macroscopic stretch in~\eqref{eq1}, while the onset of the regime is controlled by the effective stretch of the prestretched reinforcements. We therefore may consider the possibility of two distinct equilibria
(purely extensional and combined extension and torsion) \begin{equation}
\begin{array}{l}
 \lambda=\lambda_{\rm ext},\ \kappa=0,\\[2pt]
 \lambda=\lambda_{\rm ext,tor},\ \kappa=\kappa_{\rm exttor}=\hat\kappa(\lambda_{\rm ext,tor}),
\end{array}
\label{eqdef}
\end{equation}

As anticipated, differently from the classical dual Poynting effect, in the present setting no external compressive load is required to induce torsion. Humidity-induced shortening of the amorphous matrix, transmitted through matrix--fiber coupling, reduces the effective stretch of the prestretched crystalline fraction, thereby driving the reinforcements into the dual Poynting regime and selecting the coupled extension--torsion equilibrium in the supercontracted state.

\vspace{0.3 cm}
\paragraph*{Water uptake model---}
Water exchange with the environment is described by a first-order kinetic process governed by an effective diffusion coefficient $D$. For a cylindrical geometry of reference radius $R_e$, this leads to a characteristic exchange rate
\[
\beta=\frac{A D}{V}=\frac{2D}{R_e},
\]
where $A$ and $V$ denote the lateral surface area and volume of the fiber, respectively.

The environmental humidity history is recorded as time--humidity pairs $\{t_i,RH_i\}$, where the relative humidity $RH(t)\in[0,1]$ is treated as a dimensionless external driving signal. 

We model uptake as a single-rate first-order process,
\begin{equation}\label{diff}
\dot c(t)=\beta\bigl(RH(t)-c(t)\bigr),\qquad c(0)=0,
\end{equation}
equivalently written in convolution form with kernel $h(t)=\beta e^{-\beta t}$
with
\begin{equation}
c(t)=\int_0^t h(t-\tau)\,RH(\tau)\,\mathrm{d}\tau .
\end{equation}
The ODE is integrated numerically after spline interpolation of the measured signal $RH(t)$.
Here $c(t)\in[0,1]$ is the normalized internal water content.

Absorbed water irreversibly disrupts hydrogen bonds in the amorphous matrix. Following the modeling framework of Ref.~\cite{JMPSNoi}, we assume that the fraction of broken (hydrogen) bonds $m_b$ is proportional to the absorbed water content and is thus identified with $c(t)$. To enforce irreversibility, we introduce the history-dependent evolution
\[
m_b(t)=\max_{0\le\tau\le t}c(\tau),\qquad m(t)=1-m_b(t).
\]
 Here $m(t)$ measures the percentage of unbroken bonds.
\vspace{0.3 cm}

\paragraph*{Experimental comparison---}
When the ambient humidity varies in time, the internal water content becomes time dependent, $c=c(t)$, and the remodeling stretch evolves as $\lambda_m(t)=\lambda_m(c(t))$. In terms of the fraction of intact hydrogen bonds $m(t)$, and following Ref.~\cite{JMPSNoi}, we set $\lambda_m(t)=\sqrt{m(t)}$, thereby directly linking irreversible hydration-induced bond disruption to axial shortening of the amorphous matrix.

\begin{figure}[htbp]
	\begin{center}
    \centerline{\includegraphics[width=7.5 cm]{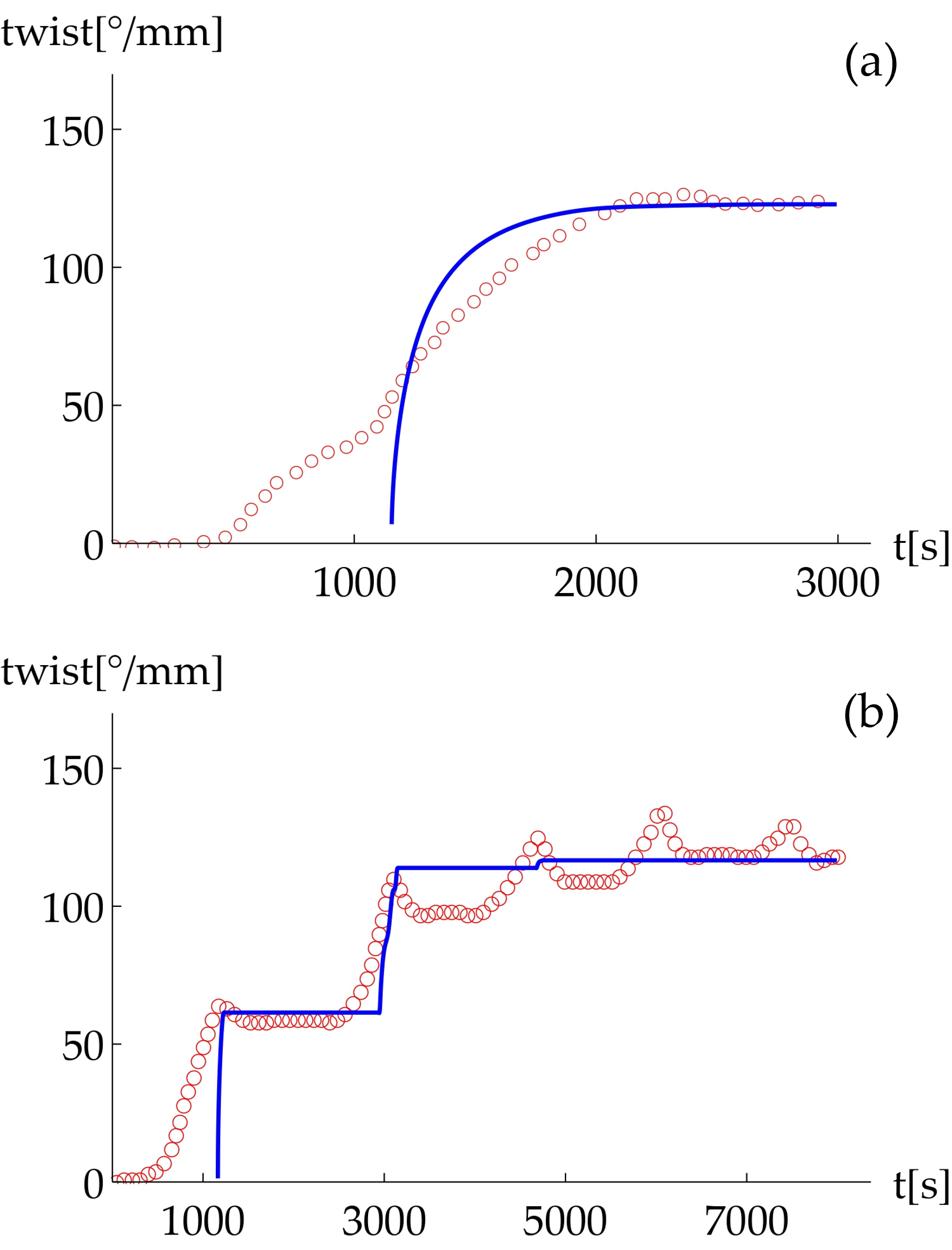}}
		\caption{Versicolor dragline silk fibers ($R_e=3.35\,\mu$m, $H=88$ mm \cite{expexp}): experiments (markers) versus model predictions (solid lines) for (a) monotonic and (b) cyclic humidity protocols. 
Model parameters: $\mu_T=50$ MPa, $E_L=200$ GPa, $\mu_L=1$ MPa, $\alpha=0.12$, $D=1.6\times10^{-14}$ m$^2$/s, $\lambda_f^{\rm res}=1.1$ (a);sm
$\mu_T=82$ MPa, $E_L=250$ GPa, $\mu_L=1$ MPa, $\alpha=0.56$, $D=1\times10^{-14}$ m$^2$/s, $\lambda_f^{\rm res}=1.1$ (b).
See End Matter for details.}
		\label{versifit}
	\end{center}
\end{figure}

With a paddle load $N_{\mathrm{paddle}}=0.1\,\mathrm{g}$ \cite{expexp}, the residual prestretch of the crystalline reinforcements is taken as $\lambda_f^{\mathrm{res}}=1.1$, consistent with the stress levels reported in Ref.~\cite{JMPSNoi}. 
Using parameter values compatible with independent multiscale estimates --such as $E_L\sim10^2$~GPa for $\beta$-sheet nanodomains \cite{cunniff1994mechanical,termonia1994molecular,fraternali2020tensegrity} and MPa-scale shear moduli for the amorphous fraction \cite{JMPSNoi} (see End Matter for a detailed discussion)-- the model quantitatively reproduces both monotonic (Fig.~\ref{versifit}a) and cyclic (Fig.~\ref{versifit}b) torsional responses. 
In both protocols, torsion emerges only beyond a critical humidity level, consistent with the well-established threshold for supercontraction.

The experiments shows a slight healing (reversible) effect, with a small rewind under decreaing humidity that is neglected in our model. 
Possible extensions accounting for partial rebonding could be straigthly incorporated within the same framework \cite{de2010damage}, but are not considered here. The small differences of the optimal material parameters required to match the two protocols are consistent with the experimental observation that the first humidity ramp in the cyclic test is not perfectly superposable to the monotonic response.\vspace{0.2 cm}

\paragraph*{Conclusions---}
We have shown that the torsional response observed in spider dragline silk during supercontraction can be explained without invoking any intrinsic geometric chirality or prescribed helical organization of the macromolecular architecture. Within a transversely isotropic nonlinear elastic framework, humidity-driven remodeling of the amorphous matrix, mechanically constrained by a prestretched crystalline fraction, naturally gives rise to a spontaneous torsional deformation as a manifestation of a dual Poynting-type mechanism.

The central result is that torsion emerges as a macroscopic mode driven by an internal mismatch of spontaneous strains, generated by hydration-induced matrix shortening relative to the reinforcing network. Unlike previous interpretations based on preassigned helical microstructures, the present theory demonstrates that axial remodeling alone, when embedded in nonlinear anisotropic elasticity, is sufficient to trigger torsion as an emergent response.

By coupling a diffusion-based description of humidity uptake with irreversible matrix remodeling and fiber plasticity, the model quantitatively reproduces both monotonic and cyclic torsional measurements~\cite{expexp} using parameter values consistent with independent multiscale estimates for spider silk. This establishes a direct and physically transparent link between molecular-scale processes (hydrogen-bond disruption and $\beta$-sheet unfolding), mesoscopic matrix--fiber interactions, and the observed macroscopic torsional actuation.

More broadly, our results identify humidity-induced torsion as a generic outcome of internal remodeling in fiber-reinforced soft materials, rather than a peculiarity of silk microstructure. The framework therefore provides a minimal and physically grounded route to environmentally driven torsional actuation, with direct implications for the design of bioinspired, humidity-responsive devices based on simple matrix--fiber architectures.

\bibliographystyle{apsrev4-2}

\bibliography{apssamp}

\section*{End Matter}

\subsection*{Model parameters and comparison with literature}

The adopted parameter values are consistent with those previously reported for spider dragline silk. In particular, $E_L=200$~GPa matches the axial modulus of the $\beta$-sheet crystalline phase obtained from computational modeling~\cite{cunniff1994mechanical}, while Termonia~\cite{termonia1994molecular} employed a value of $160$~GPa for crystalline $\beta$-sheets, also adopted in more recent microstructural models~\cite{fraternali2020tensegrity}. The MPa-scale shear moduli $\mu_T$ and $\mu_L$ are compatible with the moduli reported for the amorphous and soft components of spider silk~\cite{JMPSNoi}. 

\begin{figure}[htbp]
	\begin{center}
    \centerline{\includegraphics[width=7.5 cm]{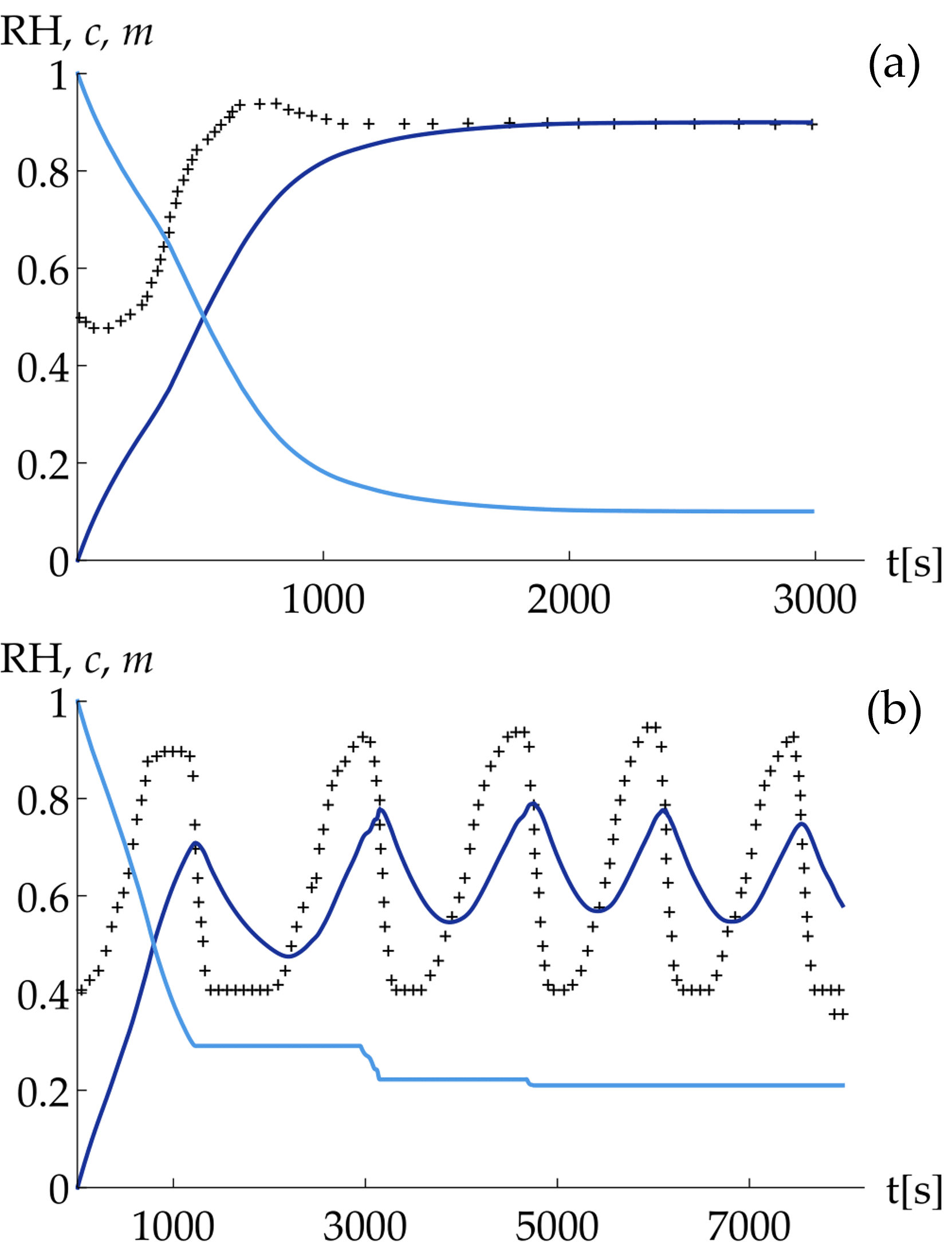}}
		\caption{\label{rhcm}Versicolor dragline silk fibers ($R_e=3.35\,\mu$m, $H=88$ mm \cite{expexp}): experimental RH (markers) and model predictions (solid lines) for the internal normalized water concentration $c$  (dark blue) and fraction of intact hydrogen bonds $m$ for (a) monotonic and (b) cyclic humidity protocols. 
Model parameters: $\mu_T=50$ MPa, $E_L=200$ GPa, $\mu_L=1$ MPa, $\alpha=0.12$, $D=1.6\times10^{-14}$ m$^2$/s, $\lambda_f^{\rm res}=1.1$ (a);
$\mu_T=82$ MPa, $E_L=250$ GPa, $\mu_L=1$ MPa, $\alpha=0.56$, $D=1\times10^{-14}$ m$^2$/s, $\lambda_f^{\rm res}=1.1$ (b).}
	\end{center}
\end{figure}

Regarding hydration kinetics, an effective diffusion coefficient $D\simeq10^{-14}\,\mathrm{m^2/s}$ yields characteristic hydration time scales of the order of minutes for micrometric fiber radii, consistent with the minute-scale humidity transients observed experimentally~\cite{expexp}. While the simple diffusion law~\eqref{diff} is sufficient to capture the main features of the response, more refined transport kinetics could further improve the quantitative accuracy, particularly in the transient regime (see Fig. \ref{versifit}). In Fig. \ref{rhcm} we show the model predictions for the internal normalized water concentration $c$ and fraction of intact hydrogen bonds $m$ togheter with the experimental RH for the monotonic and cyclic humidity protocols.

\subsection*{Linearization of the transversely isotropic strain-energy density}

We consider the incompressible transversely isotropic strain-energy density
\begin{equation}
\label{eq:W_SM}
\begin{aligned}
W(I_1,I_4,I_5)
&=\frac{\mu_T}{2}(I_1-3)
+(\mu_T-\mu_L)\,(X-X_0)\\
&\quad
+\frac{\mu_L-\mu_T}{2}\,(Y-Y_0)\\
&\quad+\frac{E_L+\mu_T-4\mu_L}{8}\,(X-X_0)^2 ,
\end{aligned}
\end{equation}
with
\begin{equation}
\label{eq:XY_SM}
\begin{array}{ll}
X=I_4+\alpha I_1, &\qquad
Y=I_5,\\
X_0=1+3\alpha,& \qquad
Y_0=1,\end{array}
\end{equation}
where $\alpha>0$ is a coupling term and $\mu_T,\mu_L,E_L$ are the transverse/longitudinal shear moduli and the
longitudinal Young modulus, respectively. The invariants are
\[
I_1=\mathrm{tr}\,\mathbf{B},\qquad
I_4=\mathbf{a}\cdot\mathbf{a},\qquad
I_5=\mathbf{a}\cdot(\mathbf{B}\mathbf{a}),
\]
where $\mathbf{B}=\mathbf{F}\mathbf{F}^T$, $\mathbf{a}=\mathbf{F}\mathbf{A}$, and $\mathbf{A}$
is the unit fiber direction in the reference configuration.

\paragraph{Small-strain expansion.}
Let
\begin{equation}
\label{eq:Fpert}
\mathbf{F}=\mathbf{I}+\epsilon\,\mathbf{H},\qquad 0<\epsilon\ll1,
\end{equation}
and define the infinitesimal strain tensor
\begin{equation}
\label{eq:eps_def}
\boldsymbol{\varepsilon}=\tfrac12(\mathbf{H}+\mathbf{H}^T).
\end{equation}
Then $\mathbf{B}=\mathbf{I}+2\epsilon\,\boldsymbol{\varepsilon}+O(\epsilon^2)$ and, to first order,
\begin{equation}
\label{eq:inv_1st}
\begin{aligned}
I_1 &= 3+2\epsilon\,\mathrm{tr}\,\boldsymbol{\varepsilon}+O(\epsilon^2),\\
I_4 &= 1+2\epsilon\,(\mathbf{A}\cdot\boldsymbol{\varepsilon}\mathbf{A})+O(\epsilon^2),\\
I_5 &= 1+4\epsilon\,(\mathbf{A}\cdot\boldsymbol{\varepsilon}\mathbf{A})+O(\epsilon^2).
\end{aligned}
\end{equation}
Consequently,
\begin{equation}
\label{eq:XY_1st}
\begin{aligned}
X-X_0 &= 2\epsilon\Big[(\mathbf{A}\cdot\boldsymbol{\varepsilon}\mathbf{A})
+\alpha\,\mathrm{tr}\,\boldsymbol{\varepsilon}\Big]+O(\epsilon^2),\\
Y-Y_0 &= 4\epsilon\,(\mathbf{A}\cdot\boldsymbol{\varepsilon}\mathbf{A})+O(\epsilon^2).
\end{aligned}
\end{equation}

\paragraph{No linear term under incompressibility.}
Set $a:=\mathbf{A}\cdot\boldsymbol{\varepsilon}\mathbf{A}$ and $t:=\mathrm{tr}\,\boldsymbol{\varepsilon}$.
Using \eqref{eq:XY_1st} in the linear terms of \eqref{eq:W_SM} (note that $(X-X_0)^2=O(\epsilon^2)$),
\begin{equation}\begin{array}{lll}
\label{eq:W_Oeps}
W&=&\epsilon\Big[\mu_T t+2(\mu_T-\mu_L)(a+\alpha t)+2(\mu_L-\mu_T)a\Big]\\
&+& O(\epsilon^2)=\epsilon\big(\mu_T+2\alpha(\mu_T-\mu_L)\big)t+O(\epsilon^2),\end{array}
\end{equation}
so the $O(\epsilon)$ fiber contributions cancel identically. For incompressible perturbations,
\begin{equation}
\label{eq:incomp_lin}
t=\mathrm{tr}\,\boldsymbol{\varepsilon}=0,
\end{equation}
hence $W=O(\epsilon^2)$.

\paragraph{Quadratic energy.}
Collecting the $O(\epsilon^2)$ terms of \eqref{eq:W_SM} (on the subspace \eqref{eq:incomp_lin})
gives
\begin{equation}\begin{array}{l}
\label{eq:Wlin_SM}
W
=\epsilon^2\Big[\mu_T\,\boldsymbol{\varepsilon}:\boldsymbol{\varepsilon}
+k\,(\mathbf{A}\cdot\boldsymbol{\varepsilon}\mathbf{A})^2\Big]+O(\epsilon^3),
\\
k=(\mu_L-\mu_T)+\frac{E_L}{2}.\end{array}
\end{equation}
Equation~\eqref{eq:Wlin_SM} coincides with the classical quadratic form of linear incompressible
transversely isotropic elasticity.

 \end{document}